\documentstyle[12pt]{article}
\def\ni{\noindent} 

\def\no{\nonumber}
\begin{document}
\begin{center}
{\Large {\bf Extended BRST invariance in topological \\Yang Mills  
theory revisited}}\\
\vspace{1cm} 

{\large Nelson R. F. Braga  and Cresus F. L. Godinho}  \\
\vspace{1cm}

 Instituto de F\'\i sica, Universidade Federal  do Rio de Janeiro,\\
Caixa Postal 68528, 21945  Rio de Janeiro,
RJ, Brazil\\

\vspace{1cm}

\end{center}

\vspace{1cm}
\abstract  

Extended BRST invariance (BRST plus anti-BRST invariances) provides in 
principle a natural way of introducing  the complete gauge fixing 
structure associated to a gauge field theory in the minimum 
representation of the algebra. However, as it happens in topological 
Yang Mills theory, not all gauge fixings can be 
obtained from a symmetrical extended BRST algebra, where antighosts 
belong to the same representation of the Lorentz group of the 
corresponding ghosts. We show here that a field redefinition makes 
it possible to start with an 
extended BRST algebra with  symmetric  ghost  antighost spectrum 
and arrive at the gauge fixing action of topological Yang Mills theory.


\vspace{1cm}

\vspace{3cm}

\noindent braga@if.ufrj.br; godinho@if.ufrj.br

\vfill\eject

It was shown by Witten  \cite{Wi,Rev} how to build up a quantum field 
theory involving the vector gauge potential whose partition functional 
generates topological invariants like the Donaldson polynomials. Soon 
after that, Baulieu and Singer\cite{BS} 
have shown that this, so called topological Yang Mills theory, can be 
obtained
by  an appropriate  gauge fixing of the topological invariant classical 
action

\bigskip
  
\begin{equation}
\label{YMA}
S_0\,=\,\int d^4x \Big( - {1\over 4} Tr ( F_{\mu\nu} {\tilde F}^{\mu\nu} )
\,\Big)
\end{equation}

\noindent where ${\tilde F}^{\mu\nu}\,\equiv \, 
{1\over 2}\epsilon^{\mu\nu\rho\sigma} 
F_{\rho\sigma} $. 

Their approach was to quantize the action (\ref{YMA}) along the standard 
BRST 
prescriptions but writing out the BRST transformations of the gauge fields 
as the sum of the BRST transformations of standard Yang Mills theory plus a 
topological ghost $\Psi^\mu\,$ that by itself removes the  degrees of 
freedom  
associated with the gauge field $\,A_\mu\,$

\begin{equation}
\label{trans}
\delta A_\mu\,=\,\Psi_\mu\,+\,D_\mu c\,.
\end{equation}

\noindent Then the field content of the theory is enlarged in order to 
gauge fix the  action,
taking into account the reducibility of  (\ref{trans}). Also the topological
Yang Mills action was obtained from a BRST quantization procedure in an
 alternative way by Labastida and Pernici\cite{LaP}.

The quantization procedure of ref. \cite{BS} is completely based on the 
sole BRST 
invariance of the gauge fixing action. The antighosts and auxiliary fields 
are introduced as trivial pairs.
They do not belong to the minimum representation of the BRST algebra.
It is known  however \cite{Extalg} that one can associate
to any classical gauge symmetry an extended BRST (BRST plus anti-BRST) 
algebra.
One of the advantages of considering an extended algebra rather than a 
standard 
one is that the antighosts and auxiliary fields enter in this formulation 
as basic ingredients  of the minimum representation of the algebra and 
not just as trivial pairs.  
The idea of introducing anti BRST transformations in topological Yang Mills 
theory 
was  discussed in refs. \cite{Ba1} , \cite{LP} and \cite{De}. An extended 
BRST algebra 
associated with this 
theory was then presented in ref. \cite{PT}. 
However, in this reference the connection with the action of Baulieu and 
Singer is only possible by means of a gauge fixing action involving terms 
with different scaling dimensions. They also use some field redefinitions 
that do not preserve  the dimension and the number of degrees of freedom. 

\medskip

The standard way of introducing the anti-BRST transformations is to define 
them in a symmetric  way with respect to the BRST ones. Following this  
approach one introduces antighosts in the same representation of the 
Lorentz group as the corresponding ghosts.   This would correspond, in 
the case of the topological Yang Mills theory to an extended 
BRST algebra that  can be written as\cite{PT}

\begin{eqnarray}
\label{Alg1}
\delta_1 A_\mu &=&  \Psi_\mu \,+\, D_\mu c \no\\
\delta_1 c &=& \phi_1\, + {1\over 2}\  [c,c] \no\\
\delta_1 \Psi_\mu  &=& - D_\mu \phi_1 \,-\,\lbrack c , 
\Psi_\mu \rbrack\no\\
\delta_1 \phi_1 &=& - \lbrack c , \phi_1 \rbrack \no\\
\delta_1 {\overline c} &=& b \no\\
\delta_1 b &=& 0  \no\\ 
\delta_1 {\overline \Psi}_\mu &=& - k_\mu \no\\
\delta_1 \lambda &=& \eta_1 \no\\
\delta_1 \phi_2 &=& \eta_2 \no\\
\delta_1 k_\mu &=& 0\no\\
\delta_1 \eta_1 &=& 0 \no\\
\delta_1 \eta_2 &=& 0 \no\\
\end{eqnarray}

\begin{eqnarray}
\label{Alg2}
\delta_2 A_\mu &=&  {\overline\Psi}_\mu \,+\, D_\mu 
{\overline c} \no\\  
\delta_2\  {\overline c} &=& \phi_2 \,-\,
{ 1 \over 2}\  \lbrack {\overline c}\,,\,{\overline c}\,] \no\\
\delta_2 \,{\overline \Psi}_\mu  &=& - D_\mu \phi_2 \,-\,\lbrack\,
{\overline c}\, , 
{\overline \Psi}_\mu \rbrack\no\\
\delta_2 \,\phi_2 &=& - \,\lbrack {\overline c}\, ,\, \phi_2 \, \rbrack 
\no\\
\delta_2 c &=& \lambda \,-\,b\,-\,[\,c\, , \,{\overline c}\, \rbrack\no\\
\delta_2 \Psi_\mu &=& D_\mu \lambda \,+\,k_\mu \,-\,[c , 
{\overline \Psi_\mu }]\,-\,
\lbrack{\overline c } , \Psi_\mu \rbrack   \no\\
\delta_2 \lambda &=&  - \, \eta_2 \,-\, \lbrack{\overline c}\,, 
\lambda\, \rbrack 
\,-\, \lbrack c , \phi_2 \rbrack \no\\
\delta_2 \phi_1 &=& - \eta_1 \,- \lbrack{\overline c} , \phi_1 
\rbrack \,-\,\lbrack c , 
\lambda^\prime \rbrack \no\\
\delta_2 \eta_2 &=& \delta_2 \Big( - [\,{\overline c} , \lambda ] - 
[ c , \phi_2 ] \Big) \no\\
\delta_2 \eta &=& \delta_2 \Big( - [\,{\overline c} , \phi ] - 
[ c , \lambda ] \Big) \no\\
\delta_2 b &=& \delta_2 \lambda - \delta_2 \Big( [ \, c , {\overline c} ]
\Big) \no\\
\delta_2 k_\mu &=& \delta_2 \Big(   D_\mu \lambda + 
[c , {\overline \Psi}_\mu ]\,+\,
[ {\overline c} , \Psi_\mu ] \Big)
\end{eqnarray}

\ni where we are representing BRST and anti-BRST transformations 
respectively as 
$\delta_1$ and $\delta_2$ and $[\,,\,]$ means a graded commutator.

\bigskip
This symmetric approach leads to a simpler algebraic structure for the 
extended algebra. 
However, in general not all gauge fixings of a gauge theory can be 
implemented by using
just this kind of symmetrical ghost antighost pairs. This is precisely
 what happens  in 
topological Yang Mills theory. In order to enforce the 
self duality condition on $F_{\mu\nu}\,$ one has to introduce a self dual 
anti symmetric 
antighost  tensor  $ {\overline \chi}^{+\mu\nu}\,$ corresponding to the 
ghost vector $\Psi^\mu$\cite{BS}. 
The  difference between the so called geometrical antighosts, like 
${\overline \Psi^\mu}$, that come from a symmetrical extended formulation 
and the actual gauge fixing antighosts,
like ${\overline \chi}^{+\mu\nu}\,$,  normally introduced as parts of
trivial BRST doublets was discussed in \cite{Ba2}. This reference 
introduces a general method for starting  with a completely symmetrical set
of geometrical ghost antighost pairs and then, by adding some trivial 
BRST doublets, arrive at a gauge fixing action corresponding to some 
particular gauge fixing.
The idea there is to  include some appropriate terms in the  action that 
will have the role of eliminating the unwanted variables from the model 
by a supersymmetric compensation in the functional integration. 
This general mechanism is called  a "transmutation" of geometrical 
antighosts into gauge fixing antighosts. 
 
We will show here that it is possible to start with the symmetric
ghost antighost set of algebras of eqs. (\ref{Alg1}) and (\ref{Alg2})
and redefine the fields in such a way that an extended algebra involving 
the gauge fixing antighost emerges.
The topological Yang Mills action is then build up from a double 
(BRST anti-BRST) variation of a gauge fixing boson. 
Let us consider the following field redefinitions:
 
\begin{eqnarray}
\label{redef}
{\overline \Psi}_\mu &=& {\overline \Psi}_\mu^{\prime}\,+\,
 D^\nu {\overline \chi}^{+}_{\mu\nu} 
\,+\,D_\mu {\overline \rho}\no\\
k_\mu &=& k_\mu^{\prime}\,+\, D^\nu b^+_{\mu\nu} \,+\,D_\mu d\,\,.
\no\\
\end{eqnarray}

\noindent Where  ${\overline \chi}^{+}_{\mu\nu} $ and $b^+_{\mu\nu} $ 
are anti symmetric self dual tensor fields  carrying three independent 
components each one. The primed fields are necessary in order to find a
representation for the extended BRST algebra at interacting level and are 
assumed to vanish when the coupling constant goes to zero.
(In other words, explicitely including the coupling constant $g$,
 omitted in the article for simplicity, we 
would have: ${\overline \Psi}_\mu \,= \,g\,{\overline \Psi}_\mu^{\prime}\,
+\, D^\nu {\overline \chi}^{+}_{\mu\nu} \,+\,D_\mu {\overline \rho}$ and 
the same kind of thing for $k_\mu$).
Equations (\ref{redef}) are just shifting
the field $ \,{\overline \Psi}_\mu\,$, separating the four components 
in three of ${\overline \chi}^{+}_{\mu\nu} $ plus one of 
${\overline \rho}$. Observe that this shift in the first of equations 
(\ref{redef}) generates a new symmetry. The new ghosts of the second 
equation  have precisely the role of fixing this symmetry.

Associated to this new set of fields we find the following extended 
algebra:

\begin{eqnarray}
\label{NAlg1}
\delta_1 A_\mu &=&  \Psi_\mu \,+\,D_\mu c \no\\
\delta_1 c &=& \phi_1\,+\,c c \no\\
\delta_1 \Psi_\mu  &=& - D_\mu \phi_1\,+\,\lbrack c , 
\Psi_\mu \rbrack \no\\
\delta_1 \phi_1 &=& - \lbrack c , \phi_1 \rbrack\no\\
\delta_1 {\overline c} &=& b \no\\
\delta_1 b &=& 0  \no\\ 
\delta_1 {\overline \chi}^{+}_{\mu\nu} &=& -  b^+_{\mu\nu} \no\\
\delta_1 \lambda &=& \eta_1 \no\\
\delta_1 \eta_1 &=&  0 \no\\
\delta_1 \phi_2 &=& \eta_2 \,-\,\lbrack \,{\overline c}\,,\,b\,
\rbrack  \no\\
\delta_1 b^+_{\mu\nu} &=& 0\no\\
\delta_1 {\overline \rho} &=& d \no\\
\delta_1 d &=& 0\no\\
\delta_1 \eta_2 &=& \lbrack\, b\,,\,b\,\rbrack \no\\
\delta_1  {\overline \Psi}_\mu^{\prime} &=& \lbrack \,\Psi^\nu \,
+\,D^\nu c\,,\,{\overline \chi}^{+}_{\mu\nu}\,\rbrack \,+\,
\lbrack \Psi_\mu \,+\,D_\mu c\,,\,{\overline \rho} \rbrack\,
-\, k_\mu^{\prime} \,-\,2 D_\mu d \no\\
\delta_1 k_\mu^{\prime} &=& \lbrack \Psi^\nu \,+\, D^\nu c 
\,,\,b^+_{\mu\nu} \rbrack \,+\,\lbrack \Psi_\mu \,+\, D_\mu c 
\,,\,d \rbrack 
\end{eqnarray}

\begin{eqnarray}
\label{NAlg2}
\delta_2 A_\mu &=&  {\overline \Psi}_\mu^{\prime}\,+\,
D^\nu {\overline \chi}^{+}_{\mu\nu},
+\,D_\mu {\overline \rho}\,+\, D_\mu {\overline c} \no\\
\delta_2\  {\overline c} &=& \phi_2 \,+\,{\overline c}{\overline c}
\no\\
\delta_2 {\overline \chi}^{+}_{\mu\nu}  &=& 
\lbrack \,{\overline \rho}\,+\, {\overline c}\,,\, 
{\overline \chi}^{+}_{\mu\nu}\rbrack \no\\
\delta_2 {\overline \rho} &=& - \phi_2 \,+\,
{\overline \rho}\,{\overline \rho}\,+\,\lbrack \,{\overline c} \,,\,
{\overline \rho} \rbrack
\no\\
\delta_2 \phi_2 &=& - \lbrack {\overline c}\,,\,\phi_2 \rbrack \no\\
\delta_2 c &=& \lambda \,-\,b\,-\,d \,+\,\lbrack c\,,\,{\overline c}
\rbrack \no\\
\delta_2 \Psi_\mu &=& k_\mu^{\prime}\,+\,D^\nu b^+_{\mu\nu}
\,+\,D_\mu d - D_\mu \lambda \,-\,\lbrack \,c\,,\, 
{\overline \Psi^\mu}\rbrack \,-\,\lbrack\, {\overline c}\,,\,
\Psi^\mu \rbrack \no\\
\delta_2 \lambda &=&  \lbrack\, c\,,\,\lambda\,+\, 
\lbrack \, c\,\,{\overline c}\,\rbrack \,\rbrack
\,-\,\lbrack {\overline\rho}\,,\,d\,\rbrack \,+\,
\lbrack \lambda\,-\,\phi_1\,-b\,-d \,-\,cc\,+\,[ \,c\,,\,{\overline c} ]
\,\,,\,\,{\overline c}\rbrack
\no\\
\delta_2 \phi_1 &=& - \eta_1 \,+\,\lbrack \phi_1\,
+\,cc\,,\,{\overline c} \rbrack \,-\,\lbrack c\,,\,
\lambda \,+\,  [ c\,,\,{\overline c} ] \,\rbrack 
\no\\
\delta_2 \eta_2 &=& 0 \no\\
\delta_2 \eta_1 &=& \lbrack c \,\,,\,\, [c\,,\,b] \,-\, 
[\phi_1\,+\,cc\,,{\overline c}] \,+\,\eta_1\,\,\rbrack \,+\,\,
\lbrack \phi_1\,+\,cc\,,\,\lambda \rbrack\no\\
&-& [ d\,,\,d ] \,+\, \lbrack \phi_1\,-\,cc\,,\,b\rbrack \,
-\, \lbrack [ c\,,\,\phi_1\,-\,cc\,]\,,\,{\overline c}\,\rbrack
 \no\\
\delta_2 b &=& - \eta_2 \no\\
\delta_2 d &=& \eta_2 
\,-\,\lbrack {\overline c} \,+\,{\overline \rho}\,,\,b\,+\,d \rbrack
\no\\
\delta_2 b^+_{\mu\nu} &=& 
-\lbrack b\,+\,d\,,\,{\overline \chi}^{+}_{\mu\nu}\rbrack \,-\,
\lbrack {\overline c} \,+\,{\overline \rho}\,,\,b^+_{\mu\nu} \rbrack\no\\
\delta_2  {\overline \Psi}_\mu^{\prime} &=& 
\lbrack D^m {\overline \chi}^{+\,\nu}_m \,+\,{\overline \Psi}^{\prime \,\nu}
 \,,\,
{\overline \chi}^{+}_{\mu\nu} \rbrack \,+\, 
\lbrack {\overline \Psi}_\mu^{\prime}\,,\,
{\overline c} \,+\,{\overline \rho} \rbrack\,
\,\,,
\end{eqnarray}

\noindent and the anti BRST transformation for $\,k_\mu^{\prime}\,$ can be 
calculated  from $\,\delta_1\delta_2 \Psi_\mu\,=\,0\,$.
Then we consider the gauge fixing of topological Yang Mills action 
(\ref{YMA}) by adding the gauge fixing action 

\begin{equation}
\label{GFA}
S_{GF}\,=\,-\,\int d^4x \,\delta_1\,\delta_2\, {1\over 2}   Tr\, 
\Big(\, A_\mu A^\mu \,+\,( {\overline c} \,-\,{\overline \rho}\,) 
\Box^{-1} \partial_\mu \Psi^\mu
\,-\,c\, ( {\overline c}\,+{\overline \rho}\,) \,\Big)\,\,,
\end{equation}

\noindent where all the terms have the appropriate scaling dimension.  
This action is different from the one in \cite{PT} that involves  the BRST 
anti-BRST variation of a sum involving terms $A^\mu A_\mu$
and $F^{\mu\nu} F_{\mu\nu} \,$ that have different scaling dimensions.
This would make sense only if one includes additional dimensionfull 
parameters.
Also our field redefinition (\ref{redef}) is different from the one 
in \cite{PT}, where ${\overline \chi}^{+\mu\nu}$  was introduced as 
the exterior derivative  of ${\overline \Psi}_\mu$, although these 
two  quantities have different scaling dimensions.
  
Using (\ref{NAlg1}) and  (\ref{NAlg2}) , we find after redefining the 
fields  as in (\ref{redef}):

\begin{eqnarray}
S_{GF} &=& \int d^4x \,Tr\, \Big( \,A_\mu k^{\prime\mu}\,+\,
A_\mu D^\mu \,(\,d\,-\,b) \,-\, A^\mu D^\nu b^+_{\mu\nu}\no\\
&-& A_\mu [ \, \Psi^\mu \,+\,D^\mu c\,,\,{\overline c}\,]\,-\,
 ( \Psi_\mu \,+\,D_\mu c ) ( {\overline \Psi}_\mu^{\prime}\,+\,
 D^\nu {\overline \chi}^{+}_{\mu\nu} 
\,+\,D_\mu {\overline \rho}\,+\,D_\mu {\overline c}\,)\no\\
&-& {1\over 2}  ({\overline c}\,-\,{\overline \rho}\,) 
\Box^{-1} \,\partial_\mu \Big( - D^\mu \eta_1 \,+\, 
[ \Psi^\mu \,+\,D^\mu c\,]\,+\,[\,c\,,\,k^\mu\,]\no\\
&+& 
[\,\phi_1 \,+\,cc\,,\, {\overline \Psi}_\mu^{\prime}\,+\,
 D^\nu {\overline \chi}^{+}_{\mu\nu} 
\,+\,D_\mu {\overline \rho}\,]\,-\, 
[ \,{\overline c}\,,\,-D^\mu \phi_1 \,+\,[\Psi^\mu \,,\,c ] \,]\,+\,
[\,b\,,\,\Psi^\mu \,]\,\Big)\no\\
&+&{1\over 2} \Big( 2 \phi_2\,+\,{\overline c} \,{\overline c}\,
-\,{\overline \rho}
{\overline \rho} \,-\,[\,{\overline c}\,,\,{\overline \rho} \,]\,\Big)
\Box^{-1} \partial_\mu \Big( -D^\mu \phi_1\,+\, [\,\Psi^\mu\,,\,c\,]\,\Big)
\,-\,  \eta_2 \Box^{-1} \partial_\mu \Psi^\mu \,
\no\\
&+& {1\over 2}\,c \,[\,{\overline c}\,+\,{\overline \rho}\,,\,b\,+\,d\,]\,+
{1\over 2}\,(\,\phi_1\,+\,cc\,) \,( \,{\overline c}\,{\overline c}\,+\,
{\overline \rho}\,{\overline \rho}\,+\,
[ \,{\overline c}\,,\,{\overline \rho}\,]\,)\,\no\\
&-& {1\over 2}
( \lambda\,-\,b\,-\,d\,)(\,b\,+\,d\,) \,+\,\eta_1 
(\,{\overline c}\,+\,{\overline \rho})\,\Big)\,
\,\,.
\end{eqnarray}

This action represents a complete gauge fixing of the topological action
(\ref{YMA}). We can now see what happens in the limit of weak coupling 
constant. In this case all the interaction terms and also the fields 
$ {\overline \Psi}_\mu^{\prime}\, 
$ and  $\, k_\mu^{\prime}\,$ vanish. In order to compare with ref\cite{BS}
we can redefine the fields:

\begin{eqnarray}
\eta_1 &\equiv& \eta \no\\
\eta_2 &\equiv& \Box {\overline \eta}\no\\
\phi_1 &\equiv& \phi \no\\
\phi_2 &\equiv& \Box {\overline \phi}\,\,.
\end{eqnarray}

\noindent the action then becomes

\begin{eqnarray}
S_{GF} &=& \int d^4x \,Tr\, \Big({\overline \chi}^{+}_{\mu\nu} 
\partial^\mu \Psi^\nu \,-\,
({\overline \rho} + {\overline c} ) \partial_\mu \Psi^\mu \,+\,
b^+_{\mu\nu}\partial^\mu A^\nu 
\,-\,{\overline \eta}\partial^\mu\Psi_\mu 
\,-\,( b + d ) \partial_\mu A^\mu\no\\
&-& \partial^\mu c \partial_\mu ( {\overline c}\,+\,{\overline \rho} )
\,+\,{\overline \phi} \Box \phi \,-\,{\overline \rho } \eta \,-\,bd\,
-\,{b^2\over 2} \,-{d^2 \over 2} \,+\,\lambda d \Big)\,\,.
\end{eqnarray}

\noindent Path integration over the fields $\lambda$ and $\eta$ leads
to functional deltas in $d$ and ${\overline \rho}$. Thus, 
integrating over this four fields we arrive at the 
total action

\begin{eqnarray}
S &=& S_{0} \,+\, \int d^4x \,Tr\, \Big({\overline \chi}^{+}_{\mu\nu} 
\partial^\mu \Psi^\nu \,-\,
{\overline c}  \partial^\mu \Psi_\mu \,+\,
b^+_{\mu\nu}\partial^\mu A^\nu 
\,-\,{\overline \eta}\partial^\mu\Psi_\mu 
\,-\, b \partial_\mu A^\mu\no\\
&-& \partial^\mu c \partial_\mu {\overline c}
\,+\,{\overline \phi} \Box \phi \,-\,{b^2\over 2} \, \Big)\,\,.
\end{eqnarray}

This is the weak coupling limit of the action found in reference 
\cite{BS}. Thus we see  
that the field redefinitions of eq. (\ref{redef}) provide  a simple 
way of separating the relevant variables, that implement the gauge 
fixing conditions (at weak coupling)

\begin{equation}
\,\,\,\,\partial^\mu A_\mu\,=\,0\,\,\,\,\,;
\,\,\,F_{\mu\nu}\,+\, {\tilde F}_{\mu\nu}\,=\,0\,\,\,\,\,;
\,\,\,\partial^\mu \Psi_\mu\,=\,0
\end{equation}

\noindent from the other unnecessary variables.  It is interesting 
to note also the way that the $\lambda$ and $\eta$ fields cancel 
just the unwanted parts of  $\,{\overline \Psi}_\mu\,$ and $\,k_\mu \,$.

The consistency of our procedure can be checked by looking at the
dimensions of the fields:
  
\begin{eqnarray}
\label{dim}
\lbrack  \,A_\mu\, \rbrack &=&  \lbrack\,c\, \rbrack  \,=\, 
\lbrack {\overline c}\, \rbrack \,
=\,  \lbrack \, {\overline \chi}^+_{\mu\nu} \,  \rbrack \,=\,\lbrack \, 
{\overline \eta}\, \rbrack \,= \,\lbrack\,{\overline \rho}\,\rbrack\,=\, 1
\nonumber\\
\,\lbrack \, \Psi^{\mu}\, \rbrack \, &=& \,\lbrack \, 
{\overline \Psi}^{\mu}\, \rbrack \,=\, \lbrack \, \phi \, \rbrack \,= 
\,2 \nonumber\\
\lbrack \, {\overline \phi} \, \rbrack &=& 0\nonumber\\
\lbrack\,\lambda\,\rbrack &=& \lbrack\,b\,\rbrack\,=\, 
\lbrack\,b^{\mu\nu}\,\rbrack\,=\,
\lbrack\, d \,\rbrack\,=\,2\nonumber\\
\lbrack \,\eta\,\rbrack &=& \lbrack\,k^\mu \,\rbrack \,=\,3
\end{eqnarray}

\medskip

Comparing our results with  those of ref.\cite{Ba2} one sees that
here the gauge fixing action is both  BRST and anti BRST exact while 
there the action  is only BRST exact. 
Also, one sees that here the antighost ${\overline \chi}^{+}_{\mu\nu}$ 
comes from a shift in ${\overline \Psi}_\mu $.

\vskip 1cm
Acknowledgements: The authors are partially supported by CAPES, CNPq., 
FINEP and FUJB (Brazilian Research Agencies).
\vfill\eject

\end{document}